\documentclass[reprint,amsmath,amssymb,aps,prl,showpacs]{revtex4-1}
\usepackage{graphicx}
\begin{document}
\preprint{preprint Number}
\title{Critical role of two-dimensional island-mediated growth on the formation of semiconductor heterointerfaces}
\author{Esperanza Luna}
	\email{luna@pdi-berlin.de}
\author{\'Alvaro Guzm\'an}
	\altaffiliation[Permanent address: ]{ISOM and Dpto. Ingenier\'ia Electr\'onica, ETSI Telecomunicaci\'on,
	Universidad Polit\'ecnica de Madrid, Spain.}
\author{Achim Trampert}
\affiliation{Paul-Drude-Institut f\"{u}r Festk\"{o}rperelektronik,
                Hausvogteiplatz $5-7$, 10117 Berlin, Germany}
\author{Gabriel \'Alvarez}
\affiliation{Departamento de F\'{\i}sica Te\'orica II, Facultad de Ciencias F\'{\i}sicas,
                Universidad Complutense, 28040 Madrid, Spain}
\date{\today}
\begin{abstract}
We experimentally demonstrate a sigmoidal variation of the composition profile across semiconductor
heterointerfaces. The wide range of material systems (III-arsenides, III-antimonides, III-V quaternary
compounds, III-nitrides) exhibiting such a profile suggests a universal behavior. We show that sigmoidal
profiles emerge from a simple model of cooperative growth mediated by two-dimensional island formation,
wherein cooperative effects are described by a specific functional dependence of the sticking coefficient
on the surface coverage. Experimental results confirm that, except in the very early stages, island growth
prevails over nucleation as the mechanism governing the interface development and ultimately determines
the sigmoidal shape of the chemical profile in these two-dimensional grown layers. In agreement with our
experimental findings, the model also predicts a minimum value of the interfacial width,
with the minimum attainable value depending on the chemical identity of the species.
\end{abstract}
\maketitle
A central goal of modern materials physics is the control of interfaces down to the atomic level.
In particular, the behavior of layered materials depends on the atomic-scale structural roughness
and chemical mixing across the interface~\cite{HY00}. Although abrupt interfaces between
conventional semiconductors (such as III-V compounds) are fabricated and element profiles across
these interfaces are obtained with atomic resolution, the relation between the layer growth processes
and the parameters governing the interface formation and evolution is not satisfactorily understood.
In this respect, there is an ongoing discussion about how interfaces can be quantitatively described
on the basis of a growth model and whether there is a minimum interface width.

Recently Hulko\emph{~et~al.}~\cite{HU08,HU09} and Luna\emph{~et~al.}~\cite{LU08,LU09,LU10}
have shown empirically that experimental concentration profiles in III-V two-dimensional~(2D)
heterostructures, e.g.~quantum wells (QW) grown by molecular beam epitaxy (MBE), can
be accurately reproduced by a sigmoidal function of the form $x(z) = x_0 /[1 + \exp(-z/L)]$.
Here, $x_0$ denotes the nominal mole fraction of one of the species, $z$ is the position
across the interface along the growth direction, and $L$ is the parameter quantifying the
interface width ($L$ is proportional to the widely reported length $W$,
over which the concentration changes from 10\% to 90\% of its plateau value).
Moreover, the accuracy of the sigmoidal fitting seems to be independent
of the experimental technique used to obtain the element distribution~\cite{HU08,LU09,MU12} and,
more interestingly, of the compound semiconductor. In this letter, we show that a sigmoidal profile
emerges from a simple model of cooperative growth with 2D island formation. Furthermore,
the use of a generalized sigmoidal expression gives a reliable and systematic quantification
of the chemical interface. It sheds light on basic aspects of the early stages of heteroepitaxial growth,
and permits to find a correlation between the profile and the interface properties in morphologically 
perfect epitaxial layers~\footnote{The layers exhibit coherent interfaces and a structural interfacial roughness 
of $\pm1\,\mathrm{ML}$}, which have been grown in the thermodynamically controlled Frank-van-der-Merwe (FM) mode~\cite{KO08}, 
not necessarily by MBE.

Experiments show that $L$ depends strongly on the combination of materials
on both sides of the interface, but that heterostructures formed by the same material combination feature the same 
value of $L$ independently of their optimized growth methods~\footnote{Optimized growth conditions
refer to 2D grown heterostructures with smooth and coherent interfaces} or substrate temperatures
$T_{\mathrm{s}}$~\footnote{For instance,
$L=1.6~\mathrm{ML}$ for $\mathrm{Al}_{0.3}\mathrm{Ga}_{0.7}\mathrm{As}/\mathrm{GaAs}$
grown at $T_{\mathrm{s}}=580\,\mathrm{C}$;
$L=1.3~\mathrm{ML}$ for $\mathrm{In}_{0.4}\mathrm{Ga}_{0.6}\mathrm{As}/\mathrm{GaAs}$
grown at $T_{\mathrm{s}}=420\,\mathrm{C}$; whereas
$L=1~\mathrm{ML}$ for $\mathrm{In}_{0.5}\mathrm{Ga}_{0.5}\mathrm{As}/\mathrm{GaSb}$
grown at $T_{\mathrm{s}}=420\,\mathrm{C}$. Furthermore, in high-quality dilute-nitride
$\mathrm{(In,Ga)(As,N)}/\mathrm{GaAs}$ QWs,
$L_\mathrm{In}=1.4-1.6\,\mathrm{ML}$ and $L_\mathrm{N}=1-1.3\,\mathrm{ML}$ regardless of
$T_{\mathrm{s}} = 375\,\mathrm{C}$ or $420\,\mathrm{C}$, even if grown at different laboratories}.
It could be argued that different $T_{\mathrm{s}}$ might modify diffusion and intermixing processes
at the interface and therefore account for the different values of $L$.
This argument, however, does not explain the different values of $L$
for dissimilar heterostructures despite their same $T_{\mathrm{s}}$ or, conversely,
that the same heterostructure features identical $L$ when grown at different $T_{\mathrm{s}}$.
Therefore, the material interface property relation seems to be more complex than a
mere increase in intermixing with increasing $T_{\mathrm{s}}$. Furthermore, the results suggest
that there is a material-specific limitation in the interface width. The wide range of materials exhibiting
a sigmoidal profile suggests a universal behavior determined by fundamental processes
occurring during growth.

A sigmoidal growth has been reported to occur in biological, geological and chemical processes
with cooperative effects~\cite{ZH03,*DU85,*GU96}, wherein the binding of one atom or molecule
affects the binding of the subsequent atoms or molecules. The experimental observation of sigmoidal
profiles at semiconductor heterointerfaces suggests that similar phenomena occur in materials science.
We infer that the species (atoms or molecules, hereafter we will use both terms indistinctly)
which are involved in the growth and interface formation process form
a strong cooperative system, where the binding affinity (quantified by the sticking coefficient
$s$) changes with the amount of atoms that have already been bound. More concretely,
we assume that the rate of adsorption of A atoms (let us consider the growth of a pseudobinary 
semiconductor
alloy $\mathrm{A}_{x_0}\mathrm{B}_{1-x_0}\mathrm{C}$ on top of a binary compound $\mathrm{BC}$) 
is proportional \emph{both} to the surface concentration of chemisorbed A atoms $\sigma$
and to the surface concentration of free A sites $(\sigma_{0}-\sigma)$, where $\sigma_{0}$
is the total surface concentration of A sites:
\begin{equation}
	\label{eq:ed1t}
	\frac{d\sigma}{d t} = {\frac{s_1 J}{\sigma_{0}}}\sigma(t)(\sigma_{0}-\sigma(t)).
\end{equation}
Here $s_1$ is a parameter that depends on the material system (we will see later that it corresponds
to the initial island growth sticking coefficient) and $J$ is the impingement rate on an area of the substrate
corresponding to one adsorption site~\cite{BE88,BE92}. If we assume that, after fast local rearrangements
not described by a simple kinetic equation, the thickness $z$ is proportional to the time variable $t$,
then the variation of the surface coverage of A atoms $\theta = \sigma/\sigma_{0}$ is described
by the differential equation:
\begin{equation}
	\label{eq:ed1}
	\frac{d\theta}{d z} = \frac{s_1 J}{r}\theta(z)(1-\theta(z)),
\end{equation}
where the constant $r$ is the mean growth rate of the pseudobinary alloy. 
The solutions of Eq.~(\ref{eq:ed1}) with initial conditions between zero and one are sigmoidal functions.
Equations~(\ref{eq:ed1t}) and~(\ref{eq:ed1}) feature a surface-coverage dependent sticking
coefficient $s(\theta)=s_1\theta(1-\theta)$. Incidentally, we mention that the experimentally
measured value of the mole fraction $x$ is simply proportional to $\theta$.
\begin{figure}
\includegraphics[width=9cm]{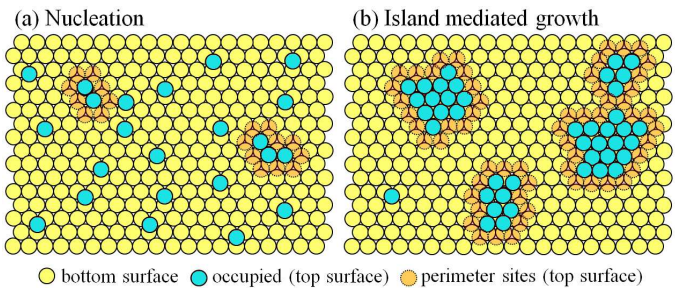}
\caption{\label{fig:eden} (Color online) Schematic location of the chemisorbed atoms on
a surface when
(a) nucleation and (b) island growth dominates. The island-mediated growth proceeds
via Eden cluster formation: in each growth step, empty perimeter sites next-neighbored
to a seed (or to an occupied site) become occupied.}
\end{figure}

Comprehensive investigations on the MBE growth of the II-VI compound $\mathrm{CdTe}$
showed that the sticking coefficients for both the $\mathrm{Cd}$ and the $\mathrm{Te}$
atoms were not constant~\cite{LI92}. Based on a remarkably good fitting of their data,
Litz\emph{~et~al.} suggested the existence of a precursor state during MBE growth, in which atoms
are physisorbed to the growing surface in the manner of Kisliuk's theory of precursor-mediated
adsorption~\cite{LI92}. In his theory, Kisliuk distinguished two types of physisorbed molecules,
intrinsic precursor molecules (ipm) and extrinsic precursor molecules (epm), which are physisorbed
above vacant and occupied chemisorption sites, respectively, albeit he assumed that adsorbed
molecules are uncorrelated~\cite{KI57}. Ensuing computer simulations revealed that molecules adsorbed
via an epm mechanism tend to aggregate into 2D clusters or islands~\cite{HO85}.
Becker\emph{~et~al.}~\cite{BE88,BE92} extended Kisliuk's model to include these lateral positional
correlations. In their model, Becker\emph{~et~al.} addressed chemisorption as a combination
of two mechanisms, each corresponding to the two types of precursor states introduced by Kisliuk:
the ipm are responsible for a seeding process in which a molecule is directly chemisorbed above
a vacant site and serves as a nucleus for the formation of an island, while epm are responsible for
island growth by initial physisorption above an occupied chemisorption site followed by lateral jumps,
until a vacant chemisorption site is reached. In this latter process, attractive adsorbate-adsorbate
interactions enhance the precursor binding potential near the island edges, so that the epm will most
likely become chemisorbed next to an already chemisorbed atom at the island edge, generating
2D compact Eden-type clusters~\cite{HO85,EV93,[{Eden growth was introduced in }] [{ as a model
for the formation of cell colonies in which each cluster grows following a simple geometrical rule:
starting from an already existing single occupied lattice site or nucleation center, empty adjacent sites are
occupied in each subsequent growth step; as a result of this mechanism Eden clusters have a compact structure.}]ED61},
as it is illustrated in ~Fig.~\ref{fig:eden}. Furthermore, Becker \emph{et~al.}\ proposed the additive contributions
of the (ipm) nucleation and (epm) island growth terms to the sticking coefficient: for the nucleation process,
they assumed a Langmuir form where the sticking coefficient is proportional to the fraction
of empty sites, i.e., to $s_0(1-\theta)$ where $s_0$ is the initial sticking coefficient~\cite{KO08},
while for the island growth term they assumed dependencies such as $\sqrt{1-\theta}$,
which do not lead to pure sigmoidal profiles.

To gain insight into the phenomena underlying the interface formation, we have generalized
Eq.~(\ref{eq:ed1}) by adding the Langmuir dependence of the seeding process (which
describes independent nucleation sites that adsorb no more than one adsorbate
and leads to a random spatial distribution of nucleation centers on the surface)
to our growth term~(\ref{eq:ed1}). The latter describes the attractive adsorbate-adsorbate
interaction previously mentioned through the $\theta(1-\theta)$ dependence.
Thus, Eq.~(\ref{eq:ed1}) is extended to:
\begin{equation}
	\label{eq:ed2}
	\frac{d\theta}{d z}
	=
	\frac{J}{r}\left[s_0(1-\theta(z))+s_1\theta(z)(1-\theta(z))\right],
\end{equation}
or, equivalently,
\begin{equation}
	\label{eq:ed3}
	\frac{d\theta}{d z}
	=
	\frac{1}{L(1+\theta_0)}(1-\theta(z))(\theta_0+\theta(z)).
\end{equation}
Equations~(\ref{eq:ed2}) and~(\ref{eq:ed3}) feature the generalized sticking coefficient
\begin{equation}
	\label{eq:s}
	s(\theta)
	=
	s_0(1-\theta)+s_1\theta(1-\theta),
\end{equation}
the fraction $\theta_0=s_0/s_1$ denotes the ratio of the initial sticking coefficients from the nucleation
and island growth terms, respectively, and
\begin{equation}
	\label{eq:l}
	L = \frac{r/J}{s_0+s_1}.
\end{equation}
We remark that this equation gives the aimed connection between the interfacial profile,
the growth process, and the material system. In particular, it gives the quantitative relation
between the interface width $L$, the kinetic parameters $r$ and $J$, and the microscopic parameters
$s_0$ and $s_1$. Furthermore, it refines the definition of an \emph{abrupt} interface.
Since $s_0+s_1\leq2$, a non-vanishing interface width is predicted for any semiconductor heterointerface.

The general solution of Eq.~(\ref{eq:ed3}) can be written in the form
\begin{equation}
	\label{eq:sol}
	\theta(z) = 1 - \frac{1+\theta_0}{1+e^{(z-z_0)/L}},
\end{equation}
where $z_0$ is an integration constant that merely shifts the $z$ coordinate. Information on the main
processes governing the growth and interface formation can be gained from inspection of Eq.~(\ref{eq:sol}).
\begin{figure}
\includegraphics[width=8cm]{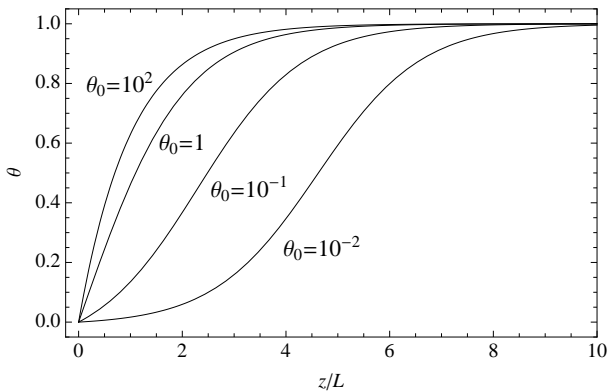}
\caption{Variation of the profiles given by Eq.~(\ref{eq:sol}) as
              $\theta_0=s_0/s_1$ ranges from $10^{-2}$ to $10^2$.
              The limit $\theta_0\gg1$ corresponds to nucleation, while the limit $\theta_0\ll1$
              corresponds to island growth.\label{fig:curves}}
\end{figure}
In Fig.~\ref{fig:curves} we plot the profiles obtained from Eq.~(\ref{eq:sol}) with $\theta(0)=0$
[which corresponds to $z_0=-L\ln(\theta_0)$] and $\theta_0$ ranging from
$10^{-2}$ to $10^2$. The asymmetry of the profiles is a quantitative measure
of the relative weight of the nucleation and the island growth terms: in the limit $\theta_0\gg1$,
i.e., $s_0\gg s_1$, where nucleation is the only mechanism, the profiles are
very asymmetric and tend to $\theta(z)=1-\exp(-z/L)$; the asymmetry of
the profiles decreases with $\theta_0$, and at $\theta_0=1$, i.e., when
$s_0=s_1$, an inflection point appears in the profile; finally, in the limit
$\theta_0\ll 1$ where island growth dominates, we recover the pure,
symmetric sigmoidal profile $\theta(z)=1/[1+\exp(-z/L)]$. Note, however,
that if $s_0=0$ the solution of the differential equation~(\ref{eq:ed1})
with $\theta(0)=0$ is identically zero, which shows that nucleation
must always exist. However, in this case, 2D island growth occurs from
the very early stages of the interface formation, since very few but already
existing nuclei are sufficient to promote further atom attachment at their rims
(Eden growth, cf.~Fig.~\ref{fig:eden}). Obviously this process is initially
less efficient than seeding, and initially it takes more time to increase the
composition profile to significative values.

\begingroup
\squeezetable
\begin{table*}
\caption{Parameters obtained from nonlinear fittings of the experimental profiles of 2D
              morphologically perfect III-V semiconductor heterostructures to the theoretical
              profile of Eq.~(\ref{eq:sol}). Growth methods: molecular beam epitaxy (MBE),
              solid phase epitaxy (SPE), metal-organic chemical vapor deposition (MOCVD).
              References are given for data taken from the bibliography. As a measure of the
              overall quality of the fit we include in the last column the coefficient of
              determination $R^2$.
              \label{tab:data}}
\begin{ruledtabular}
	\begin{tabular}{ccccccc}
	Material & Ref. & Growth method & $x_0$ (\%) & $\theta_0$ & $L~\mathrm{(ML)}$ & $R^2$\\
	\hline\\
	$\mathrm{Al}_{x_0}\mathrm{Ga}_{1-x_0}\mathrm{As}/\mathrm{GaAs}$ & & MBE &
	$29.46\pm0.02$ & $(1.7\pm0.8)\times10^{-3}$ & $1.65\pm0.02$ & 0.9997\\
	$\mathrm{In}_{x_0}\mathrm{Ga}_{1-x_0}\mathrm{As}/\mathrm{GaAs}$ & & MBE &
	$39.8\pm0.06$ & $(5.5\pm0.7)\times10^{-3}$ & $1.31\pm0.01$ & 0.9996\\
	$\mathrm{In}_{x_0}\mathrm{Ga}_{1-x_0}\mathrm{As}/\mathrm{GaAs}$ & & SPE &
	$34.2\pm0.02$ & $(2.3\pm1.8)\times10^{-3}$ & $1.61\pm0.03$ & 0.9991\\
	$\mathrm{In}_{x_0}\mathrm{Ga}_{1-x_0}\mathrm{As}/\mathrm{GaSb}$ & \cite{TA10} & MBE &
	$49.2\pm2$ &  & $1.0\pm0.1$\\
	$\mathrm{In}_{1-x_0}\mathrm{Ga}_{x_0}\mathrm{Sb}/\mathrm{InAs}$ & \cite{MA06} & MBE &
	$81\pm3$ & $(1\pm1)\times10^{-2}$ & $1.3\pm0.1$ & 0.9970\\
	$(\mathrm{In},\mathrm{Ga})(\mathrm{N},\mathrm{As})/\mathrm{GaAs}$ & & MBE &
	(In) $32.6\pm0.2$ & $(2.1\pm2.2)\times10^{-3}$ & $1.62\pm0.03$ & 0.9990\\
	 &  & &
	(N) $5.7\pm0.1$ & $(1.5\pm0.7)\times10^{-2}$ & $1.3\pm0.1$ & 0.9892\\
	$(\mathrm{In},\mathrm{Ga})(\mathrm{N},\mathrm{As})/\mathrm{GaAs}$ & \cite{LU08,CH04} & MBE &
	(In) $34.4\pm0.1$ & $(2.9\pm2.5)\times10^{-3}$ & $1.45\pm0.03$ & 0.9996\\
	 &  & &
	(N) $4.33\pm0.06$ & $(6.7\pm5.4)\times10^{-3}$ & $1.0\pm0.1$ & 0.9988\\
	$\mathrm{In}_{x_0}\mathrm{Ga}_{1-x_0}\mathrm{N}/\mathrm{GaN}$ & & MOCVD &
	$23.9\pm0.2$ & $(1.8\pm0.4)\times10^{-2}$ & $1.5\pm0.1$ & 0.9986\\
	$\mathrm{In}_{x_0}\mathrm{Ga}_{1-x_0}\mathrm{N}/\mathrm{GaN}$ & \cite{PR11} & MOCVD &
	$7.15\pm0.02$ & $(1\pm1)\times10^{-3}$ & $(0.177\pm0.002)\mbox{ nm}$ & 0.9995\\
	\end{tabular}
\end{ruledtabular}
\end{table*}
\endgroup

\begin{figure*}
\includegraphics[width=18cm]{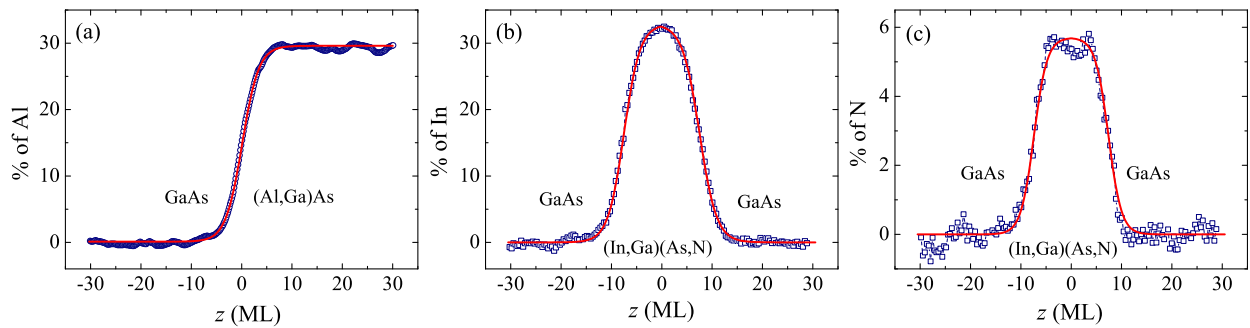}
\caption{\label{fig:abc} (Color online) Experimental composition profile (open symbols) and sigmoidal
function fits (solid line) across (a) an interface of (Al,Ga)As/GaAs. (b) and (c) represent the In and N profiles,
respectively, across the interfaces in the QW structure (In,Ga)(As,N)/GaAs. The experimental relative
error of the data points is typically $1.5\%$ for Al and In and $4.6\%$ for N.
(The growth direction goes from left to right).}
\end{figure*}

We have analyzed experimental element profiles using the functional form in Eq.~(\ref{eq:sol}) and derived
$\theta_0$ and $L$ for different III-V semiconductor heterostructures. We have investigated more than 60 samples
comprising 2D morphologically perfect III-arsenides, III-antimonides, quaternary III-V alloys 
and III-nitrides heterostructures, which were fabricated using different methods.
High-resolution and analytical transmission electron microscopy (TEM) techniques have been used
to quantitatively determine the change in stoichiometry across the interfaces.
Representative results are summarized in Table~\ref{tab:data}. As observed, in all cases, regardless
of the materials system and of the growth method, the largest values for $\theta_0$ are
$\theta_0\approx2\times10^{-2}$, which after due account of the accuracy, are consistent with a pure 
sigmoidal appearance of the experimental curves. Heterostructures formed by the same material combination feature the same $L$ value. The slight difference in $L$ for $\mathrm{(In,Ga)As}/\mathrm{GaAs}$ interfaces formed by MBE and solid phase epitaxy (SPE), respectively, arises from experimental uncertainties at SPE. Despite the complexity in the growth and composition determination of $\mathrm{(In,Ga)(As,N)}/\mathrm{GaAs}$ interfaces, high-quality heterostructures exhibit
$L_\mathrm{In}=1.4-1.6\,\mathrm{ML}$ and $L_\mathrm{N}=1-1.3\,\mathrm{ML}$, even if grown at different laboratories. Additionally,
$L$ varies with the combination of materials at both sides of the interface. As an example of these data and of the resulting fits, we show in Fig.~\ref{fig:abc}(a) an experimental
profile across an $\mathrm{(Al,Ga)As}/\mathrm{GaAs}$ interface with $L = 1.6\,\mathrm{ML}$, while
Figs.~\ref{fig:abc}(b) and ~\ref{fig:abc}(c) correspond to the In and N distributions, respectively, across an
$\mathrm{(In,Ga)(As,N)}/\mathrm{GaAs}$ QW. In this case, the fit yields
$L(\mathrm{In}) = 1.6\,\mathrm{ML}$ and $L(\mathrm{N}) = 1.3\,\mathrm{ML}$
for the $\mathrm{(In,Ga)(As,N)}/\mathrm{GaAs}$ interface, while for the
$\mathrm{GaAs}/\mathrm{(In,Ga)(As,N)}$ interface we obtain
$L(\mathrm{In}) = 1.7\,\mathrm{ML}$ and $L(\mathrm{N}) = 1.3\,\mathrm{ML}$  (cf.~Table~\ref{tab:data}).
In all our experiments, we have not found neither significant asymmetric distribution with an exponential-like
behavior nor curves reflecting $\theta_0\approx 1$. Furthermore, the experimental data are remarkably
well reproduced by sigmoidal profiles, like those represented in Fig.~\ref{fig:abc}.
We thus conclude that, in spite of the ubiquitous nucleation term, the epm-mediated 2D island growth
mechanism is the primary process at the forming interfaces and determines the general shape of the
transition zone at the heterointerface. Note that, although the exact adsorption processes
occurring in the epitaxy of semiconductors are generally unknown~\cite{KR02}, it is very well established
by reflection high-energy electron diffraction or scanning tunneling microscopy measurements
that in the FM growth mode the completion of the layers occurs in an island-mediated 2D process~\cite{HE04}.

Furthermore, our comparative analysis of several III-V heterostructures reveals a limitation in the interface width,
which is specific to each material system, in agreement with the predicted limit in the value of $L$. As shown above,
the interface width $L$ contains information both on the growth kinetics and on the adsorbate-surface relation
(through $s_0$ and $s_1$). In general, the magnitude of the sticking coefficient and the adsorption dynamics are
ultimately determined by the exact details of the potential energy surface (PES) that describes the particular
molecular-surface interaction and is specific to a given molecule-surface system~\cite{KO08}. Our experimental
findings suggest that, for samples grown under optimized conditions~\cite{Note2}, the minimum attainable interface
width is not kinetically driven, but thermodynamically driven through the adsorbate-surface interaction. The experimental results
support this hypothesis. The width $L$ at the $\mathrm{A}_{x_0}\mathrm{B}_{1-x_0}\mathrm{C}/\mathrm{BC}$
interface would be determined ultimately by the $\mathrm{A}_\mathrm{adsorbate}$ - $\mathrm{BC}_\mathrm{surface}$ PES,
which depends on the chemical identity of the surface and adsorbing atoms.
Hence, the different values of $L$ for $\mathrm{(Al,Ga)As}/\mathrm{GaAs}$ compared to $\mathrm{(In,Ga)As}/\mathrm{GaAs}$
(cf. ~{Table I}) may arise from the different PES in the
$\mathrm{Al}_\mathrm{adsorbate}$ - $\mathrm{GaAs}_\mathrm{surface}$ and
$\mathrm{In}_\mathrm{adsorbate}$ - $\mathrm{GaAs}_\mathrm{surface}$ systems, respectively; and,
similarly, the $\mathrm{In}_\mathrm{adsorbate}$ - $\mathrm{GaAs}_\mathrm{surface}$ interaction will differ
from the $\mathrm{In}_\mathrm{adsorbate}$ - $\mathrm{GaSb}_\mathrm{surface}$ one,
therefore leading to distinct values of $L$ in the $\mathrm{(In,Ga)As}/\mathrm{GaAs}$ and $\mathrm{(In,Ga)As}/\mathrm{GaSb}$
material systems, as experimentally observed. Finally, the intriguing identical $L$ for high-quality
$\mathrm{(In,Ga)(As,N)}/\mathrm{GaAs}$ QWs regardless of the growth procedure or laboratory
is also explained because the PES would be similar in all cases. Equation~(\ref{eq:l}) also predicts a dependence
on the substrate orientation because the PES (and hence $s_0$ and $s_1$) depends on the crystallographic
orientation of the surface above which growth proceeds.

Finally, we would like to point out that there is a close relation between Eq.~(\ref{eq:ed2}) and a similar equation obtained
in the context of silicon oxidation~\cite{SU99}, where a sigmoidal behavior was also reported.
Although seemingly dealing with a different problem, it has been recently shown that the initial stages of Si(001)
oxidation are determined by a combination of a Langmuir-type adsorption and 2D island growth~\cite{SU99,OH08}
in close analogy with our study. Moreover, sigmoidal laws have been reported in the context of nanocrystals
growth in solution~\cite{BI08}, where recent investigations reveal a complex growth dynamics involving a 2D nucleation
and growth process via atomic attachment from the solution phase~\cite{ZH09,LI10}.

We conclude that a sigmoidal variation of the composition across an interface
(or, in general, of any order parameter defining a system) directly arises from this specific growth
dynamics characterized by strong cooperative interactions with 2D island formation and, therefore,
it has a general character and is not restricted to MBE or MOCVD grown semiconductor heterointerfaces.
We show that the prevalence of precursor-mediated 2D island growth over nucleation is the primary
mechanism governing the interface development in 2D semiconductor heterostructures and responsible
for the sigmoidal shape of the chemical profile at the interface. Moreover, while kinetics factors confirm
as valuable control parameters for adjusting the interface width during the fabrication process,
there is a minimum attainable interfacial width, which is dictated by the molecule-surface
interaction potential and is exclusive for each material system considered, since the latter depends
on the chemical identity of the species involved.

We would like to acknowledge C.~Van~de~Walle (UCSB), K.~H.~Ploog and H.~Riechert for fruitful discussions
and S.~F\"{o}lsch for a critical reading of the manuscript.
%
\end{document}